\documentclass[final,3p]{elsarticle}

\usepackage{amsthm,amsmath,amssymb}
\usepackage{mathrsfs}
\usepackage{booktabs}
\usepackage{comment}
\usepackage{framed,multirow}
\usepackage{microtype}
\usepackage{wrapfig}

\usepackage{lineno}
\usepackage[colorlinks=true,breaklinks=true,pdftex]{hyperref}
\modulolinenumbers[1]

\journal{ }

\biboptions{authoryear}

\begin{document}

\begin{frontmatter}

\title{Point normal orientation and surface reconstruction by incorporating isovalue constraints to Poisson equation}

% USE FOR THE REVIEW PROCESS

\author[1,3]{Dong Xiao}
\ead{xiaod18@mails.tsinghua.edu.cn}

\author[2,4]{Zuoqiang Shi\corref{cor}}
\ead{zqshi@tsinghua.edu.cn}

\author[1,3]{Siyu Li}
\ead{lisiyu21@mails.tsinghua.edu.cn}

\author[5]{Bailin Deng}
\ead{DengB3@cardiff.ac.uk}

\author[1,3]{Bin Wang\corref{cor2}}
\ead{wangbins@tsinghua.edu.cn}

\cortext[cor]{Corresponding author at: Yau Mathematical Sciences Center, Tsinghua University, Beijing, China}
\cortext[cor2]{Corresponding author at: School of Software, Tsinghua University, Beijing, China}
%Note: There are two corresponding authors, Bin Wang and Zuoqiang Shi. If this journal allows only one corresponding author, keep Bin Wang.

\address[1]{School of Software, Tsinghua University, Beijing, China}
\address[2]{Yau Mathematical Sciences Center, Tsinghua University, Beijing, China}
\address[3]{Beijing National Research Center for Information Science and Technology, Beijing, China}
\address[4]{Yanqi Lake Beijing Institute of Mathematical Sciences and Applications, Beijing, China}
\address[5]{School of Computer Science and Informatics, Cardiff University, Wales, UK}

\begin{abstract}
Oriented normals are common pre-requisites for many geometric algorithms based on point clouds, such as Poisson surface reconstruction. However, it is not trivial to obtain a consistent orientation. In this work, we bridge orientation and reconstruction in the implicit space and propose a novel approach to orient point cloud normals by incorporating isovalue constraints to the Poisson equation. In implicit surface reconstruction, the reconstructed shape is represented as an isosurface of an implicit function defined in the ambient space. Therefore, when such a surface is reconstructed from a set of sample points, the implicit function values at the points should be close to the isovalue corresponding to the surface. Based on this observation and the Poisson equation, we propose an optimization formulation that combines isovalue constraints with local consistency requirements for normals. We optimize normals and implicit functions simultaneously and solve for a globally consistent orientation. Thanks to the sparsity of the linear system, our method can work on an average laptop with reasonable computational time. Experiments show that our method can achieve high performance in non-uniform and noisy data and manage varying sampling densities, artifacts, multiple connected components, and nested surfaces. The source code is available at \url{https://github.com/Submanifold/IsoConstraints}.
\end{abstract}


\end{frontmatter}

% Comment out for final accepted paper submission

\section{Introduction}\label{sec:introduction}
Point sets with consistently oriented normals are pre-requisites for many applications in computer graphics, such as modeling, rendering, and reconstruction~\citep{DBLP:journals/cgf/SchertlerSG17, DBLP:journals/tog/MetzerHZGPC21}. Although it is easy to obtain unoriented normals via local plane or surface fitting~\citep{LIII1901On, 2005Jets}, globally consistent orientation\textemdash{}where all normals point to the same side of the surface\textemdash{}cannot be completely determined locally. This makes normal orientation a challenging task in geometric modeling.

Many traditional orientation approaches focus on local consistency, i.e., adjacent points should have similar orientations~\citep{DBLP:conf/siggraph/HoppeDDMS92, 2003Piecewise, DBLP:conf/vmv/KonigG09}. This is typically achieved by propagating the normal orientation along a minimum spanning tree. However, the propagation process can be easily affected by noise or sharp features. Moreover, local errors may be propagated to large areas due to the lack of global supervision. 
%Therefore, local consistency alone does not guarantee robust orientation. 
\cite{DBLP:journals/vc/WangYC12} and \cite{DBLP:journals/cgf/SchertlerSG17} formulate the orientation process as an energy optimization problem. Although the minimization is globally optimal, the constraints are mainly dominated by local consistency. When the orientation of a region is inconsistent with the surrounding regions, the energy function only increases near the boundary of this region. Therefore, local errors may still cause large-scale inconsistent orientation. Given that local consistency is usually not sufficient to guarantee robustness, some approaches consider other types of conditions. \cite{DBLP:journals/tog/KatzTB07} check the visibility of points and achieve view-dependent orientation and reconstruction. 
\cite{DBLP:journals/tog/MetzerHZGPC21} propose a dipole propagation scheme and orient point patches in a global electric dipole field. Robust global orientation remains a non-trival problem with room for improvements.

The problem of implicit unoriented reconstruction, where an implicit surface is reconstructed from a point cloud without normals, is also closely related to normal orientation. 
It is pointed out in~\cite{DBLP:journals/cgf/MullenGDCA10} that the two tasks are of nearly the same difficulties. Some remarkable unoriented reconstruction approaches have appeared in recent years such as PGR~\citep{2022pgr} and iPSR~\citep{2022ipsr}. In this work, we mainly focus on bridging orientation and reconstruction in the implicit space.

In implicit surface reconstruction, the surface is expressed as the isosurface of an implicit function in the ambient space. For consistency between the extracted surface and the point cloud shape, the function values at the sample points should be close to the isovalue corresponding to the surface. We call this property the \textit{isovalue constraints}. In this work, we incorporate isovalue constraints to the Poisson equation and propose a new optimization approach that solves for normals and the implicit function in one linear equation.

The Poisson equation gives the relationship between the implicit function and the vector field generated by the point normals~\citep{DBLP:conf/sgp/KazhdanBH06, DBLP:journals/tog/KazhdanH13}. Therefore, we aim to search for a globally consistent normal orientation so that the sample points satisfy the isovalue constraints. Given a set of sample point positions, we optimize their normals and the implicit function values simultaneously with an appropriate boundary condition. We minimize a target function that enforces the isovalue constraints, the Poisson equation, as well as the local consistency of normals. To numerically solve the optimization, we discretize the Poisson equation and transform the minimization into a sparse linear least squares problem. To the best of our knowledge, this is the first work that optimizes both the implicit function and normals in a single target function on the Poisson reconstruction framework.

We conduct experiments on a variety of datasets with both synthetic and real scanned point clouds. The results show that our method can achieve competitive performance on a common laptop CPU, and produce robust global orientation on datasets with different sampling densities, noise levels, thin structures, and sharp features. It is also suitable for surfaces with nested structures and multiple connected components. Moreover, we demonstrate that our approach can be effectively extended to large-scale data with millions of points by globally subsampling representative point sets and orienting the dense point cloud normals based on the implicit field generated by the reference set.

%
%\subsubsection{Subsubsection Heading Here}
%Subsubsection text here.

\section{Related works}\label{sec:related_work}

Although unoriented normals of point clouds can be directly estimated by local fitting using PCA~\citep{LIII1901On} or Jets~\citep{2005Jets}, it is difficult to determine whether the normal is pointing inward or outward of the surface via local properties only~\citep{DBLP:journals/tog/MetzerHZGPC21}. Hence, deriving a consistent orientation of point cloud normals is a challenging problem. In this section, we review several existing techniques on normal orientation and implicit unoriented reconstruction. The techniques are classified into two categories: (1) orientation based on local consistency, and (2) orientation dominated by other factors. When classifying a method into the second category, we do not mean completely ignoring local properties, but mean applying additional supervision to ensure normal consistency. Unoriented reconstruction approaches are generally classified into the second category. 

\subsection{Orientation based on local consistency}\label{sec:propagation}

The main idea of local consistency-based approaches is to propagate the known orientations to neighboring points. The seminal work in~\cite{DBLP:conf/siggraph/HoppeDDMS92} constructs a minimum spanning tree (MST) based on the similarity between adjacent normals and propagates the orientations along the tree. However, noise and sharp features may cause incorrect propagation paths. In addition, local errors may be propagated to larger areas and cause severe degradation of the performance. Although several improved flip criterion methods have been proposed~\citep{2003Piecewise, DBLP:conf/vmv/KonigG09}, their robustness is still unsatisfactory for varying inputs. Some methods formulate orientation as a global optimization problem instead. \cite{DBLP:journals/vc/WangYC12} minimize a combination of the Dirichlet energy and the coupled-orthogonality deviation to ensure that the normals are consistent with the adjacent points and perpendicular to the surface. \cite{DBLP:journals/cgf/SchertlerSG17} formulate normal propagation as a graph-based energy minimization problem and solve it by quadratic pseudo-Boolean optimization (QPBO)~\citep{1991Network}. \cite{DBLP:journals/cgf/JakobBG19} propose a parallel greedy solver on the GPU for graph-based minimization. The objective functions of these methods are still dominated by local consistency items. When inconsistent orientation occurs in a region, the energy function only increases near the region boundary. Thus, local errors may still cause large-scale inconsistent orientation due to the lack of global supervision. Moreover, these approaches may fail when orienting point clouds with multiple connected components or nested structures.

\subsection{Implicit unoriented reconstruction and orientation based on other factors}\label{sec:volume}

Since local consistency is usually not sufficient to obtain robust orientation for varying inputs, some works apply other strategies to determine orientation. \cite{DBLP:conf/sma/MelloVT03} construct a signed distance function defined on a simplicial decomposition of a bounding box, where the sign of the distance value indicates the in/out information with respect to the input point set.  \cite{DBLP:conf/visualization/XieMQ04} cluster the point set into singly-oriented groups and determine the global orientation by the active contours. \cite{DBLP:journals/tog/KatzTB07} utilize a visibility-based heuristic to recognize outer surfaces and generate view-dependent orientation and reconstruction. \cite{DBLP:conf/sgp/AlliezCTD07} build a tensor field from the Voronoi diagram and solve for an implicit function whose gradient aligns with the principal axes of the field. \cite{DBLP:journals/cgf/MullenGDCA10} obtain a signed distance function from an unsigned distance approximation by ray shooting and sign propagation. 

In recent years, researchers have made considerable progress in implicit unoriented reconstruction. VIPSS~\citep{2019VIPSS} sets up gradient norm constraints of the signed distance function and solves it by Duchon’s energy. This method achieves good performance on sparse samples and 3D sketches. However, the cubic time and space complexity restrict its application on larger-scale problems. PGR~\citep{2022pgr} regards normals and surface elements in the Gauss formula as unknowns and optimizes the parametric function space. 
%The Gauss formula of~\cite{2022pgr} and the dipole field of~\cite{DBLP:journals/tog/MetzerHZGPC21} are both related to the winding number~\citep{2018Winding}. 
%PGR is the most relevant baseline for us. 
However, it involves a dense linear system that needs to be solved on the GPU to achieve reasonable efficiency, which makes it impractical for large-scale problems. 
%However, it always requires high computing resources and relies on GPU. 
By contrast, our method only needs to solve a sparse linear system and can work on a common laptop CPU with high orientation accuracy and scalability. iPSR~\citep{2022ipsr} runs Poisson reconstruction in an iterative manner and updates normals using the generated surface of the last iteration. Although it exhibits outstanding performance and generates high-fidelity results, it currently lacks a theoretical guarantee for convergence. Different from iPSR, our method establishes an optimization problem in the implicit function space, which can be solved using a single sparse linear system. Our method also performs well on thin structures with noise.

In addition to traditional approaches, deep neural networks have been applied to gather information of different scales for orientation and reconstruction~\citep{DBLP:journals/cgf/GuerreroKOM18, 2019DeepSDF, DBLP:conf/eccv/ErlerGOMW20, DBLP:journals/cg/XiaoLSW22, DBLP:journals/cad/WangLLLC22}. Among them, \cite{DBLP:journals/cad/WangLLLC22} utilize local patches for unoriented normal estimation and apply the global features for consistent orientation. However, these methods usually require substantial training data. Some works learn the implicit function directly from the input point cloud and eliminate the need for training data~\citep{DBLP:conf/cvpr/AtzmonL20, DBLP:conf/icml/GroppYHAL20, DBLP:conf/icml/MaHLZ21, DBLP:journals/corr/abs-2109-04398, DBLP:conf/cvpr/Ben-ShabatKG22}. However, they need to operate an individual network for each shape. \cite{2021ShapeAsPoints} propose a differentiable Poisson solver for both optimization-based and learning-based unoriented reconstruction. Different from other approaches, the point positions are also optimized in their work. \cite{DBLP:journals/tog/MetzerHZGPC21} apply deep networks for local patch orientation and propose a dipole propagation strategy across the coherent patches to achieve global consistency. However, the patch partition strategy may affect the robustness of this method, especially on thin structures.

\section{Method}\label{sec:method}
In this section, we present our optimization formulation that bridges orientation and reconstruction in the implicit function space. We treat normals as variables and optimize them such that the isosurface of the implicit function is consistent with the sample points, i.e., the function values at the sample points are close to the isovalue, which forms the isovalue constraints. Our method incorporates the isovalue constraints to the discretized Poisson equation, which forms the target function together with the local consistency energy. The resulting formulation optimizes the implicit function and the point normals simultaneously, which can be solved using a sparse linear least squares system.

\subsection{Discretizing the Poisson equation}\label{sec:preliminaries}
To formulate our optimization for both the implicit function and the normals, we need to first establish a relationship between them.
In this work, we model their correlation using a discretized Poisson equation similar to the one used in Poisson surface reconstruction (PSR)~\citep{DBLP:conf/sgp/KazhdanBH06, DBLP:journals/tog/KazhdanH13}.
%To build the constraints, we need a relationship between the normals and the implicit function. In this work, we apply the Poisson equation to establish the implicit space and discretize it by the finite element method in an octree. The way we discretize the Poisson equation is similar to Poisson surface reconstruction (PSR)~\citep{DBLP:conf/sgp/KazhdanBH06, DBLP:journals/tog/KazhdanH13}, except that normals are treated as variables. 
In the following, we first provide a brief review of PSR as preliminaries. 

Given the oriented sample points $\mathcal{S} = \{(p_i, n_i)_{i=1}^{N}\}$ where $p_i \in \mathbb{R}^3$ is the point position and $n_i \in \mathbb{R}^3$ is the normal, PSR first builds an octree $\mathcal{O}$ from the point cloud. Then a vector field $\vec{V}$ is computed from the normals as  %(see Section 4 of \cite{DBLP:conf/sgp/KazhdanBH06}):  
%and establishes a vector field based on the normals. For any position $q \in \mathbb{R}^3$, the value of the vector field at point $q$ is calculated as (see Section 4 of \cite{DBLP:conf/sgp/KazhdanBH06}):
\begin{equation}
\label{equation1}
 \vec{V}(q) = \sum_{s_i=(p_i, n_i) \in \mathcal{S}} \tilde{F}_{s_i}(p_i, q) \cdot n_i
\end{equation}
with
\begin{equation}
\label{equation_fpq}
\tilde{F}_{s_i}(p_i, q) =  {\frac{1}{W(p_i)} \sum_{o \in \mathcal{N}_{D}(s_i)}{\alpha_{o,s_i} F_{o}(q) }},
\end{equation}
where $\mathcal{N}_{D}(s_i)$ denotes the set of neighbor nodes for $s_i$ in the octree of the same depth $D$. $\alpha_{o,s_i}$ is the coefficient for trilinear interpolation. $F_{o}(q)$ is an approximated Gaussian smooth function, whose value decreases when $o$ and $q$ are far apart. $W(p_i)$ is related to the local sampling density. For more details, please refer to Section 4 of \cite{DBLP:conf/sgp/KazhdanBH06}. In general, the vector field of a point $q \in \mathbb{R}^3$ is dominated by the normals of its nearby sample points. In the inset figure below, we illustrate the vector field generated by a set of samples. Here the red points represent the sample points, whereas the blue arrows indicate  the generated vector field.

\begin{wrapfigure}{r}{0.35\textwidth}
    \centering
    \includegraphics[width=1.0\linewidth]{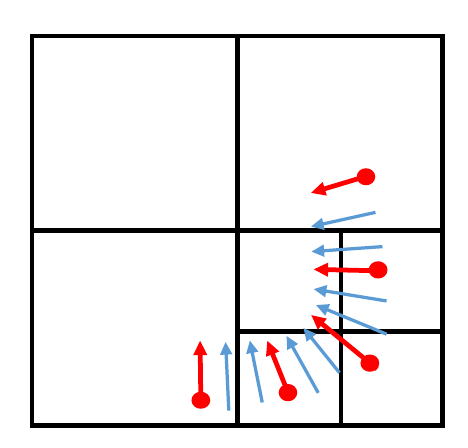}
    %%\caption{\textcolor[rgb]{1,0,0}{Illustration of the vector field.}}
    %%\label{Figure:field}
\end{wrapfigure}

We can express Equation~\ref{equation1} in a matrix form by concatenating the normals into a vector $n = (n_1^T, n_2^T, \ldots, n_N^T)^T \in \mathbb{R}^{3N}$, such that there exists a $3 \times 3N$ matrix $\mathscr{F}(q)$ where 
\begin{equation}
\label{equation_vq}
\vec{V}(q) = \mathscr{F}(q) n.
\end{equation}
We can ignore the small entries of $\mathscr{F}(q)$ due to the locality of the Gaussian function, so that $\mathscr{F}(q)$  becomes a sparse matrix. Equation~\ref{equation_vq} indicates that the vector field can be expressed as a linear function of the point normals.

The indicator function $\chi(q)$ is a typical implicit function with values of 0 outside and 1 inside the surface. By aligning the gradient of the indicator function $\chi(q)$ with the vector field $\vec{V}$, the indicator function and the vector field should satisfy the Poisson equation
%The relationship between the indicator function $\chi(x)$ and the vector field $\vec{V}$ can be expressed by the Poisson equation
\begin{equation}
\label{equation_poisson}
\Delta \chi = \nabla \cdot \vec{V}.
\end{equation}
In PSR, the indicator function is represented as a linear combination of a set of compactly supported B-spline basis functions, where each octnode $o \in \mathcal{O}$ is assigned with a basis at the node center, i.e.,
 \begin{equation}
 \label{equation_chi}
 \chi(q) = \sum_{o \in \mathcal{O}}{x_o \mathscr{B}_o(q)},
 \end{equation}
where $\{x_o\}$ are the interpolation coefficients to be solved, and $\mathscr{B}_o(q)$ is a three-dimensional B-spline basis function of degree two centered at $o$. In addition, the Poisson equation~\ref{equation_poisson} is also projected onto a series of basis functions. This results in a sparse linear system
\begin{equation}
A x = b,
\label{eq:PSRPoissonEq}
\end{equation}
where $x$ concatenates the coefficients $\{x_o\}$ for the basis functions, and the matrix $A$ and the vector $b$ has components 
\begin{equation}
 A_{ij} = <\nabla \mathscr{B}_{i}, \nabla \mathscr{B}_{j}>_{[0,1]^3}, \quad b_{i} = <\nabla \mathscr{B}_{i}, \vec{V} >_{[0,1]^3}.
\end{equation}
%is the inner product of the B-spline gradients, $b$ is related to the vector field and constructed by $\vec{V}$. Their components are $A_{ij} = <\nabla \mathscr{B}_{i}, \nabla \mathscr{B}_{j}>_{[0,1]^3}, b_{i} = <\nabla \mathscr{B}_{i}, \vec{V} >_{[0,1]^3}$. $x$ is a vector of $x_o$s and represents the interpolation coefficients. For details of the discretization process, please refer to Section 3 of~\cite{DBLP:journals/tog/KazhdanH13}. 
For details of the projection process, please refer to Section 3 of~\cite{DBLP:journals/tog/KazhdanH13}.

Different from PSR which solves the Equation~\ref{eq:PSRPoissonEq} to obtain the coefficients $x$ for the indicator function, we treat both $x$ and the normals $n$ as unknowns and use Equation~\ref{eq:PSRPoissonEq} to derive their relation. To this end, we note that the vector field $\vec{V}$ is represented as a linear function of the normals in Equation~\ref{equation_vq}. Therefore, 
\begin{equation}
b_{i} = <\nabla \mathscr{B}_{i}, \vec{V}> = \int_{[0,1]^3}{(\nabla \mathscr{B}_{i}^T(q) \vec{V}(q)) \ dq} = [\int_{[0,1]^3}{(\nabla \mathscr{B}_{i}^T(q) \mathscr{F}(q)) \ dq}] \ n.
\end{equation}
Using this condition, we can write the vector $b$ as a linear function of $n$:
\begin{equation}
b = Bn,
\end{equation}
where $B$ is a sparse matrix due to the locality of $\mathscr{B}_{i}$ and $\mathscr{F}$. It follows that the coefficients $x$ and the normals $n$ should satisfy the relation
 \begin{equation}
 \label{equation_pq}
 Ax=Bn.
 \end{equation}
This condition will be used to formulate the objective function of our optimization problem. 

%The matrices $A$ and $B$ can be named as discrete Laplacian and divergence operators, respectively~\citep{DBLP:journals/corr/abs-2206-15236}. The process of obtaining $Ax=Bn$ is similar to the discretization process for $Ax=b$ in PSR, except that $n$ is treated as variables and is placed outside in the equation.

%Given that normals are known for PSR, the implicit function is obtained by directly solving $Ax=b$. However, our method focuses on unoriented reconstruction. Normals are unknown variables in our problem. Hence, we require additional constraints to optimize both $x$ and $n$. Equation~\ref{equation_pq} gives the relationship that $x$ and $n$ need to satisfy during optimization.

\subsection{Formulation of the objective function}\label{sec:contraints}

In this section, we formulate the objective function of our optimization for the indicator function coefficients $x$ and the point normals $n$ by enforcing a set of constraints that they should satisfy. 
%We continue the notation of the previous section: $x$ represents the interpolation coefficients of the indicator function, and $n$ is the vector of point normals. They are both variables to be optimized. The objective function of our method is composed of two parts: isovalue constraints and local consistency energy.

\subsubsection{Poisson equation constraints}\label{sec:Poisson_contraints}
In the previous subsection, we use the Poisson equation to derive the relation~\ref{equation_pq} between the the indicator function and the normals. As we treat both the coefficients $x$ and the normals $n$ as variable, the condition~\ref{equation_pq} needs to be enforced by our optimization. To this end, we introduce the following term into the objective function to penalize the violation of the condition:
%Poisson equation gives the relationship between the implicit field and the normals. Thus, it should be satisfied between $x$ and $n$ during optimization. In Section~\ref{sec:preliminaries}, the Poisson equation is discretized into $Ax=Bn$ in Equation~\ref{equation_pq}. Therefore, we establish energy $E_{Poi} (x, n)$ as follows:
\begin{equation}
\label{equation7}
E_{Poi}(x, n) = ||Ax-Bn||^2.
\end{equation}
%In this manner, we incorporate isovalue contraints to the Poisson equation.

\subsubsection{Isovalue constraints}\label{sec:isovalue_contraints}
In implicit reconstruction, the reconstructed surface corresponds to an isosurface of the indicator function.
%is usually obtained by isosurface extraction, e.g., the Marching Cubes algorithm~\citep{DBLP:conf/siggraph/LorensenC87}. 
Given a point cloud that is sampled from the surface, the indicator function values at the sample points should be close to the isovalue of the surface. In this way, the extracted isosurface is consistent with the point shape. Therefore, we enforce such isovalue constraints in our optimization. Our goal is to make the indicator function equal to 0 outside the surface and 1 inside the surface. If the normals are well oriented, then the indicator function value changes rapidly from 0 to 1 near the surface. Thus, we choose the isovalue to be the midpoint $\frac{1}{2}$, and use it as the indicator function value at the sample points. Then the isovalue constraints can be written as
 \begin{equation}
\chi(p_i) = \sum_{o \in \mathcal{O}}{x_o \mathscr{B}_o(p_i)} = \frac{1}{2}, \quad i=1,2,...,N.
 \end{equation}
This can be written in a matrix form 
 \begin{equation}
Ux = \frac{1}{2} \vec{1},
 \end{equation}
where  the matrix $U \in \mathbb{R}^{N \times |\mathcal{O}|}$ has coefficients $u_{ij}= \mathscr{B}_j(p_i)$ ($i$ is index of the sample point and $j$ is index of the sorted octnode $o \in \mathcal{O}$), and  $\vec{1}$  is a vector whose components are all 1s. 
Note that the B-spline basis functions are compactly supported, $U$ is a sparse matrix. Then we enforce the isovalue constraints via the following term in the objective function:
\begin{equation}
\label{equation2}
E_{iso} (x) = ||Ux - \frac{1}{2}\vec{1}||^2.
\end{equation}

\emph{Remark.} In screened Poisson surface reconstruction (SPSR)~\citep{DBLP:journals/tog/KazhdanH13}, point constraints are introduced in their formulation to fit the isovalue. The major difference between the isovalue constraints in method and the point constraints in SPSR is that SPSR requires point normals as input, while our method treats normals as variables. Therefore, the isovalue constraints of our approach can be spread to normals during optimization with the bridging of Equation~\ref{equation7}.

The isovalue constraints and the aforementioned Poisson equation constraints in Section~\ref{sec:Poisson_contraints} are basic terms of our formulation. They allow us to search for a globally consistent normal orientation such that the indicator function values of Poisson reconstruction are equal to the isovalue on the sample points. 

\subsubsection{Local consistency constraints}\label{sec:local_contraints}

Our formulation also enforces the consistency between normals at adjacent points. Inspired by~\cite{DBLP:journals/vc/WangYC12}, we adopt a Dirichlet energy term $E_{D}(n)$ to penalize the difference between adjacent normals, and a coupled-orthogonality deviation term $E_{COD}(n)$ to enforce the orthogonality condition between the normal vectors and the underlying surface.
First, we construct a graph $\mathcal{E}$ for the sample points using their $k$-nearset neighbors (KNN), where two points $i$ and $j$ are connected with an edge when $i$ is in the KNN of $j$ or $j$ is in the KNN of $i$ (we set $k=10$ in our experiments). Using this graph, we define the terms $E_{D}(n)$ and $E_{COD}(n)$ as follows:

\begin{align}
E_{D} (n) & = \frac{1}{2} \sum_{i}{\sum_{j \in \mathcal{N}(i)}{w_{ij}||n_i - n_j||^2}},\label{equation3}\\
E_{COD} (n) &= \sum_{i}{\sum_{j \in \mathcal{N}(i)}{w_{ij}[(\frac{p_j - p_i}{||p_j - p_i||})^T (n_i + n_j)]^2 }}\nonumber \\
      &= \sum_{i}{\sum_{j \in \mathcal{N}(i)}{w_{ij}[(n_i + n_j)^T \frac{(p_j - p_i) (p_j - p_i)^T}{||p_j - p_i||^2}}(n_i + n_j)]}\label{equation4},
\end{align}
where
\begin{equation}
w_{ij} = \exp\left(-\frac{||p_i - p_j||}{\rho^2}\right),\quad  \rho = \max_{(p_i, p_j) \in \mathcal{E}}{\frac{||p_i - p_j||}{2}}.
 \end{equation}
The rationale for the term $E_{COD}(n)$ can be explained as follows. Suppose $p_i$ and $p_j$ are two nearby points on a quadratic surface patch and their geodesic midpoint is the center of the surface patch (see the inset figure below). Then their normal vectors $n_i$ and $n_j$ should satisfy~\citep{Rusinkiewicz2019}
\begin{equation}
(\frac{p_j - p_i}{\|p_j - p_i\|})^T (n_i + n_j) = 0.
\label{eq:Orthogonality}
\end{equation}
\begin{wrapfigure}{r}{0.35\textwidth}
   \centering
    \includegraphics[width=1.0\linewidth]{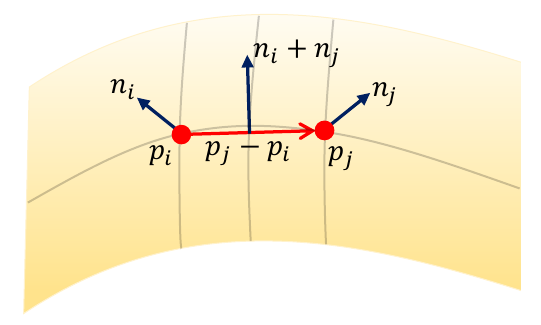}
\end{wrapfigure}
Since a smooth surface can be locally approximated around each point using a quadratic height-field surface over the tangent plane at the point, two nearby points on the surface and their normals should satisfy condition~\ref{eq:Orthogonality} approximately. Thus we use the term $E_{COD}(n)$ to enforce this condition.

We introduce the local consistency energy $E_{loc}(n)$ as the sum of $E_{D}(n)$ and $E_{COD}(n)$. Similar to~\cite{2022pgr}, we apply the L2 regularizer to ensure the stability of optimization and avoid oscillation especially for noisy inputs. Therefore,
\begin{equation}
\label{equation_loc}
E_{loc}(n) = E_{D} (n) + E_{COD} (n) + n^T n.
\end{equation}
From Equation~\ref{equation3} and~\ref{equation4}, we notice that $E_D(n)$ and $E_{COD}(n)$ are quadratic functions of $n = (n_1^T, n_2^T, \ldots, n_N^T)^T \in \mathbb{R}^{3N}$. Thus, there exists a matrix $M$ such that
\begin{equation}
\label{equation_m}
 E_{loc}(n) = n^T M n.
\end{equation}
$M$ is sparse due to the locality of nearest neighbors. 

Local consistency constraints mainly act as a regularizer and do not dominate the optimization. Decent reconstructions can also be achieved without the local consistency energy. However, enforcing local consistency is beneficial for our method especially for noisy inputs. We will show that they improves the reconstruction quality both qualitatively and quantitatively in the experiments.

\subsubsection{Boundary condition}\label{sec:overall}
Combining Equation~\ref{equation7},~\ref{equation2} and~\ref{equation_m}, the overall objective function $E(x, n)$ can be written as
\begin{equation}
\label{equation8}
    E(x, n) = E_{iso} (x) + \alpha E_{Poi}(x, n) + \beta E_{loc}(n),
\end{equation}
where $\alpha$ and $\beta$ are weights specified by the user. During experiments, the parameter $\alpha$ keeps unchanged. $\beta$ differs for noisy and noise-free inputs and is relatively small, mainly to maintain stability and avoid oscillation. The parameters do not need to be adjusted frequently.

We want the indicator function to be 0 outside the surface, $\frac{1}{2}$ on the surface, and 1 inside the surface. Therefore, the overall objective function $E(x, n)$ in Equation~\ref{equation8} must be minimized subject to a Dirichlet boundary condition, that is, the indicator value on the bounding box should be 0. Otherwise, a trivial solution $\chi(x)=\frac{1}{2}, \forall x \in \mathbb{R}^3$ and $n = \vec{0}$ exists. We set the bounding box to be slightly larger than the original object. Let $\mathcal{B}$ be the octnodes of the border, we force the coefficients of the basis functions at $\mathcal{B}$ to be 0, that is
\begin{equation}
\label{equation9}
\begin{aligned}
min_{x, n} ~& E(x, n) \\
\text{s.t.} ~& x_i = 0, i \in \mathcal{B}.
\end{aligned}
\end{equation}
According to the Poisson equation, the variation of the indicator function is mainly caused by the vector field formed by the point normals. As shown in the inset figure below, if the indicator function equals to 0 at the bounding box and 0.5 at the surface (represented by the green outline), then we can predict the normal directions to be vertical to the circle and pointing inward according to the variation of the indicator values. The indicator function transitions from 0 to 0.5, similar to ``surrounding'' the surface from the bounding box. Therefore, our method is good at identifying the inward (outward) normals of the outermost surfaces, which is helpful for noisy thin structures.

\begin{wrapfigure}{r}{0.4\textwidth}
  \centering
  \includegraphics[width=1.0\linewidth]{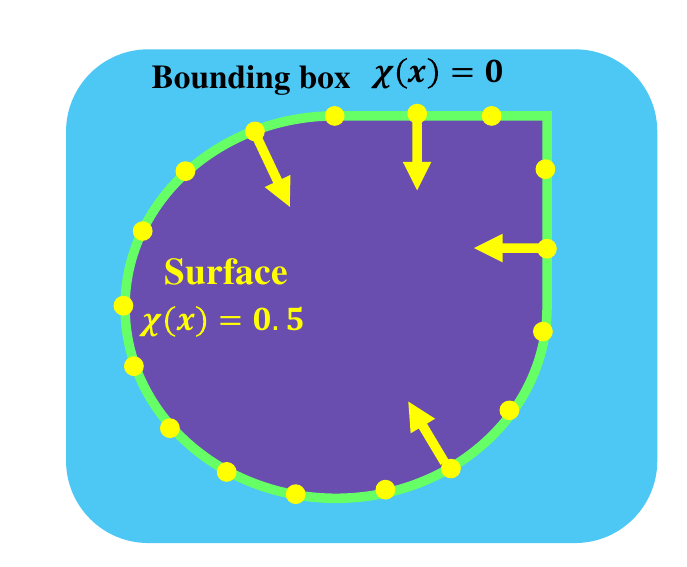}
\end{wrapfigure}

\subsection{Solving}\label{sec:solving}
From Equation~\ref{equation8}, the objective function can be expressed as follows:
\begin{equation}
  \label{equation-Exn2}
  E(x, n) = ||Ux - \frac{1}{2}\vec{1}||^2 + \alpha ||Ax-Bn||^2 + \beta (n^T M n).
\end{equation}
The boundary condition $x_i = 0, i \in \mathcal{B}$ can be enforced by crossing out the boundary components of $x$ and the corresponding columns of matrices $A$ and $U$ in the equation. This optimization problem can then be transformed into an unconstrained least squares problem by $\nabla_{x} E(x, n) =0$ and $\nabla_{n} E(x, n) =0$. It is equivalent to solving the linear system
\begin{equation}
\tilde{A}\tilde{x} = \tilde{b},
\end{equation}
where
\begin{equation}
\tilde{A}=
\begin{pmatrix}
    U^T U + \alpha A^T A & -\alpha A^T B\\
    -\alpha B^T A & \alpha B^T B + \beta M\\
\end{pmatrix},
\qquad
  \tilde{x} =
\begin{pmatrix}
    x\\
    n\\
 \end{pmatrix},
 \qquad
\tilde{b} =
\begin{pmatrix}
    U^T \frac{1}{2} \vec{1}\\
    0\\
\end{pmatrix}.
\end{equation}

We solve this linear system by the conjugate gradient method (CG)~\citep{1952CG}. $U, A, B$ and $M$ are all sparse matrices. It is worth noting that we do not explicitly calculate the matrix $\tilde{A}$ during the solving process because $U^T U, A^T A$ and $B^T B$ are not necessarily sparse. However, each step of the conjugate gradient algorithm can be written as the product of a matrix and a vector, or the inner product of two vectors. For instance, $U^T U x = (U^T(Ux)), x^T U^T Ux = (Ux, Ux)$. Therefore, the sparsity of the matrices can be maintained during the solving process. For more details, please refer to~\ref{appendixA}.
%Notice that:

%\begin{equation}
%  \label{equation_10}
%  \tilde{A}\tilde{x} =
%\begin{pmatrix}
%   U^T U x + \alpha A^T A x - \alpha A^T B n\\
%   - \alpha B^T A x + \alpha B^T B n + \beta (M + I) n\\
%\end{pmatrix}.
%\end{equation}
%Moreover, $U^T U x = U^T(Ux)$ and $x^T U^T U x = (Ux, Ux)$.

Let $|\mathcal{O}|$ be the number of octnodes, the size of matrix $A$ is $|\mathcal{O}| \times |\mathcal{O}|$ and dominates the main time and space resources during calculation since $|\mathcal{O}|$ is always larger than the point number $N$. Fortunately, the matrix is sparse with nonzero element $O(|\mathcal{O}| log(|\mathcal{O}|))$. The reason is that the construction of $A$ is similar to that of PSR. Thus, the number of its nonzero elements is equivalent to the complexity of executing the Algorithm 1 of~\cite{DBLP:journals/tog/KazhdanH13}. The logarithm occurs due to the octree depth. For $B$ and $U$, each sample point is only related to the nearby nodes and corresponding basis functions at each depth. Therefore, they will not exhibit higher complexity. When conducting multiplication of a sparse matrix and a vector, the time complexity mainly depends on the number of nonzero elements of the matrix. The time complexity of the optimization process is $O(m|\mathcal{O}| log(|\mathcal{O}|))$, where $m$ is the number of conjugate iterations.

\subsection{Orientation and reconstruction}\label{sec:recon}
Let $(\hat{x}, \hat{n})$ be the solution of Equation~\ref{equation9}, due to memory and time limitation, we set the max tree depth to be $7$ during optimization and perform 300 CG iterations. This may be not enough for high-fidelity reconstruction, but is sufficient for deciding the orientation of normals in most cases. Therefore, we do not apply the optimized indicator coefficients $\hat{x}$ for reconstruction. Instead, we apply the optimized normals $\hat{n}$ for orientation and input the oriented normals to screened Poisson surface reconstruction (SPSR)~\citep{DBLP:journals/tog/KazhdanH13} with max depth $10$. SPSR uses a conforming cascadic solver depth by depth without solving the large linear system including the octnodes of all depths.

To conduct normal orientation, we treat the optimized normals $\hat{n}$ as references. For each point, we estimate its unoriented normal (no inside/outside information) $\tilde{n}_i$ by Jets~\citep{2005Jets}. We also have an optimized normal $\hat{n}_i$ from Equation~\ref{equation9}. We set the final normal prediction of this point to $\tilde{n}_i$ if $\tilde{n}_i \cdot {\hat{n}}_{i} \ge 0$ and $-\tilde{n}_i$ otherwise. In this manner, we can obtain the inward normals\textemdash{}the normals of the outermost surface are pointing inward. For most cases, using oriented Jets normals (i.e., Jets normals oriented by our method) results in slightly higher accuracy than directly using the optimized $\hat{n}$ because Jets method mainly focuses on local properties and usually generates accurate unoriented normals. However, in cases when the local surface fitting is inaccurate such as the extremely sparse 3D sketch inputs (the first two examples of Figure~\ref{fig:Figure8}), we apply the optimized $\hat{n}$ (after normalization) as the normal prediction.

Our approach can support per-point optimization (i.e., feeding all the point normals for optimization) for more than $100$K points in common devices. For large-scale point clouds (entailing millions of points), we can subsample a representative point set from the dense point cloud and orient the whole point cloud using the implicit field generated by the representative set. When the representative set is well oriented by our method, we run SPSR to generate the implicit function of the set. The gradient field of the implicit function can be calculated by $\nabla \chi(q) = \sum_{o \in \mathcal{O}}{x_o \nabla \mathscr{B}_o(q)}$. For each point $\tilde{p}_j$ in the dense point cloud with unoriented Jets normal $\tilde{n}_j$, we calculate $\nabla \chi(\tilde{p}_j) \cdot \tilde{n}_j$ to determine the normal orientation (flip the normal if the inner product is lower than 0). This implicit orientation method is simple but effective, as the optimization of Equation~\ref{equation9} is dominated by isovalue constraints and does not rely on density sampling when the basic shape is included. In most cases, such a scale is sufficient to contain the entire basic shape and decide the normal orientation of the dense point cloud. Then, a high-fidelity reconstruction can be achieved by the oriented dense point cloud. In addition, excessive many variables are not beneficial to carry out optimization. The use of representative sets can relax the complexity of our optimization. The representative set approach is mainly inspired by~\cite{DBLP:journals/tog/MetzerHZGPC21}. The difference is that~\cite{DBLP:journals/tog/MetzerHZGPC21} utilize the dipole implicit field, whereas our approach applies the Poisson implicit field.

\section{Experiments}\label{sec:experiments}
We conduct various qualitative and quantitative experiments on orientation accuracy and reconstruction quality to demonstrate the efficacy of our approach. The experiments are performed on different datasets with varying noise, sampling densities, and artifacts.

\subsection{Comparisons}\label{sec:compare}
We compare our method with representative and state-of-the-art orientation and unoriented reconstruction approaches. For the dataset, we use the benchmark data complied by a recent survey paper~\citep{2022Survey} that collects a large number of shapes from different repositories such as 3DNet~\citep{3DNet2012}, ABC~\citep{2019ABC}, Thingi10K~\citep{2016Thingi10K}, and 3D Scans~\citep{20213DScans}. The shapes are classified as simple, ordinary, or complex according to the complexity of the shape topology and the difficulties for reconstruction. 

\begin{figure}[htb]
  \centering
  \includegraphics[width=1.0\linewidth]{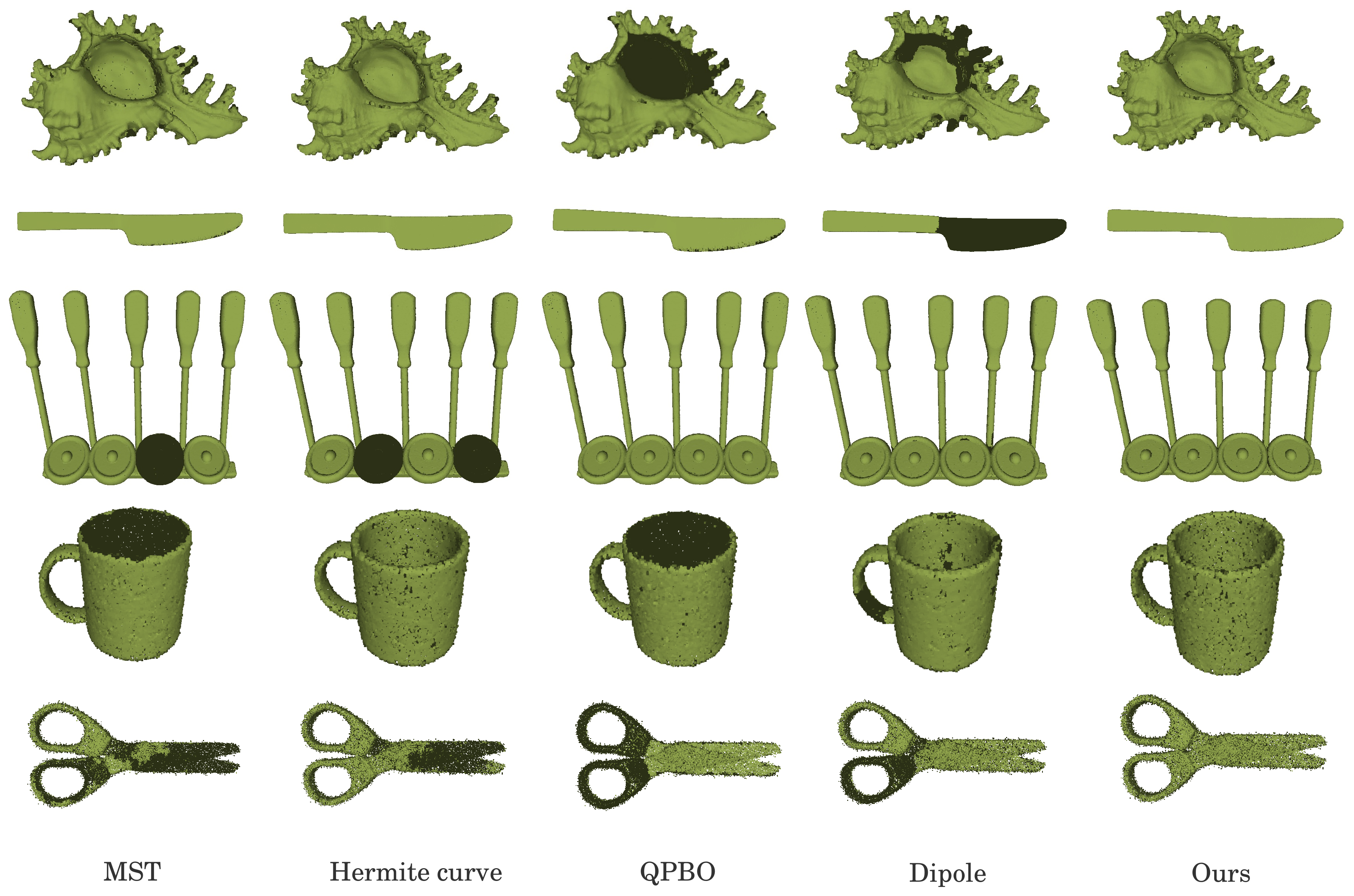}
  \caption{
           Qualitative comparisons with  MST~\citep{DBLP:conf/siggraph/HoppeDDMS92}, Hermite curve~\citep{DBLP:conf/vmv/KonigG09}, QPBO~\citep{DBLP:journals/cgf/SchertlerSG17} and Dipole~\citep{DBLP:journals/tog/MetzerHZGPC21} in terms of orientation accuracy. Our method achieves high performance in point clouds with complex topology, thin structures, and noise.} 
    \label{fig:Figure3}
\end{figure}

We carry out our experiments on two different sampling styles: 1. non-uniform but nearly no noise and 2. varying noise. For non-uniform sampling, we choose the complex shapes (total 162 shapes) and directly apply the point clouds provided by the benchmark~\citep{2022Survey} with $160$K points per shape. They are scanned by the Blensor simulator~\citep{2011Blensor}. In our approach, we use a representative set of $\frac{1}{3}$ original points (about $53$K) for optimization and orient the whole point cloud according to the reference set. We set the parameters to $\alpha = 10^4$ and $\beta = 10^{-4}$ for no-noise inputs.

For var-noise sampling, we use the ordinary shapes (total 486 shapes) and directly sample $40$K points from the surface by trimesh~\citep{2019trimesh}. To obtain varying noise, we add randomized Gaussian noise with varying noise ratio $[0.5, 1.0]$ and varying noise standard derivation $[0.002, 0.01]$ for each shape. The noise ratio represents the proportion of noisy points among all points and is different for each shape. We conduct per-point optimization (no representative set is required) for var-noise sampling and set parameters to $\alpha = 10^4$ and $\beta = 10^{-2}$. The parameter settings are maintained throughout the experiment.

During qualitative comparison, we also select several data from~\cite{DBLP:journals/tog/BergerLNTS13}, some 3D sketches from~\cite{2019VIPSS} and some real scanned point clouds from the aforementioned benchmark~\citep{2022Survey}. 

First, we compare our results with four orientation methods MST~\citep{DBLP:conf/siggraph/HoppeDDMS92}, Hermite curve~\citep{DBLP:conf/vmv/KonigG09}, QPBO~\citep{DBLP:journals/cgf/SchertlerSG17} and Dipole~\citep{DBLP:journals/tog/MetzerHZGPC21} in terms of orientation accuracy, which indicates the proportion of correctly oriented normals (where the dot product with the ground truth normal is greater than zero) among all sample points. The unoriented normals are all predicted by Jets~\citep{2005Jets}. For non-uniform sampling, the ground truth normal of a point is taken as the surface normal of the ground truth mesh where the point is located. For noisy sampling, the ground truth normal of a point is taken as the normal when it is sampled. Table~\ref{table1} shows the quantitative comparisons. In addition to average accuracy, we also count the proportion of shapes whose accuracy is greater than a certain threshold for each method ($97\%$ for complex non-uniform dataset and $90\%$ for ordinary var-noise dataset). For instance, our approach achieves orientation accuracy higher than $97\%$ in $96.3\%$ $(156/162)$ of the shapes in the complex non-uniform dataset. The results indicate that our method achieves high performance. Local consistency-based methods (MST, Hermite curve and QPBO) may fail in orienting point clouds with multiple connected components in one process. Therefore, these shapes are removed when calculating the accuracy of them.

\begin{table}[t]
\centering
\caption{Quantitative comparisons with MST~\citep{DBLP:conf/siggraph/HoppeDDMS92}, Hermite curve~\citep{DBLP:conf/vmv/KonigG09}, QPBO~\citep{DBLP:journals/cgf/SchertlerSG17} and Dipole~\citep{DBLP:journals/tog/MetzerHZGPC21} in terms of orientation accuracy.}
\label{table1}
\begin{tabular}{ccccc}
\hline

\multirow{2}{*}{Method} & \multicolumn{2}{c}{Complex non-uniform} & \multicolumn{2}{c}{Ordinary var-noise}\cr
%\cline{2-3} \cline{4-5}
\cmidrule(lr){2-3} \cmidrule(lr){4-5}
  & Avg.  & Acc.$>97\%$ & Avg. & Acc.$>90\%$  \\
\hline

MST  & $97.1\%$ & $87.2\%$ & $86.4\%$ & $66.6\%$ \\

Hermite curve  & $98.6\%$ & $92.3\%$ & $90.2\%$ & $76.6\%$ \\

QPBO  & $98.4\%$ & $89.1\%$ & $87.2\%$ & $67.9\%$ \\
%\hline

Dipole  & $95.8\%$ & $77.8\%$ & $90.1\%$ & $73.0\%$ \\
%\hline

Ours  & $\textbf{99.3\%}$  & $\textbf{96.3\%}$ & $\textbf{93.7\%}$ & $\textbf{85.2\%}$\\
\hline

\end{tabular}
\label{table_MAP}
\end{table}

  \begin{figure}[htb]
  \centering
  \includegraphics[width=0.7\linewidth]{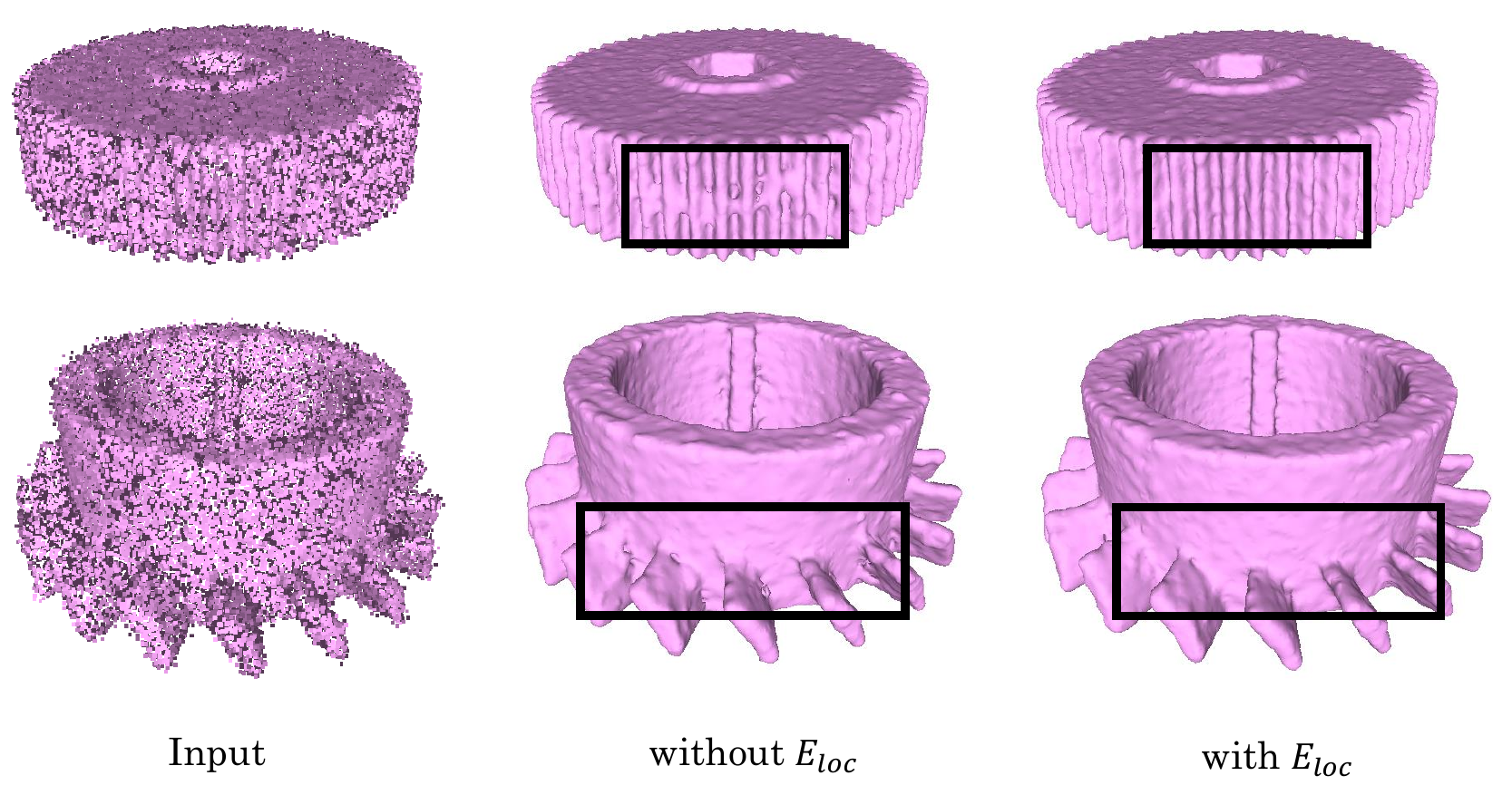}
  \caption{
           The reconstruction results with and without the local consistency energy $E_{loc}$ of two noisy inputs. We use black rectangles to emphasize the differences. It can be seen that $E_{loc}$ brings about better orientation consistency and improves the reconstruction quality near the zigzag region of the input shapes. }
           
    \label{fig:Figure_local}
\end{figure}

\begin{figure}[htb]
  \centering
  \includegraphics[width=1.0\linewidth]{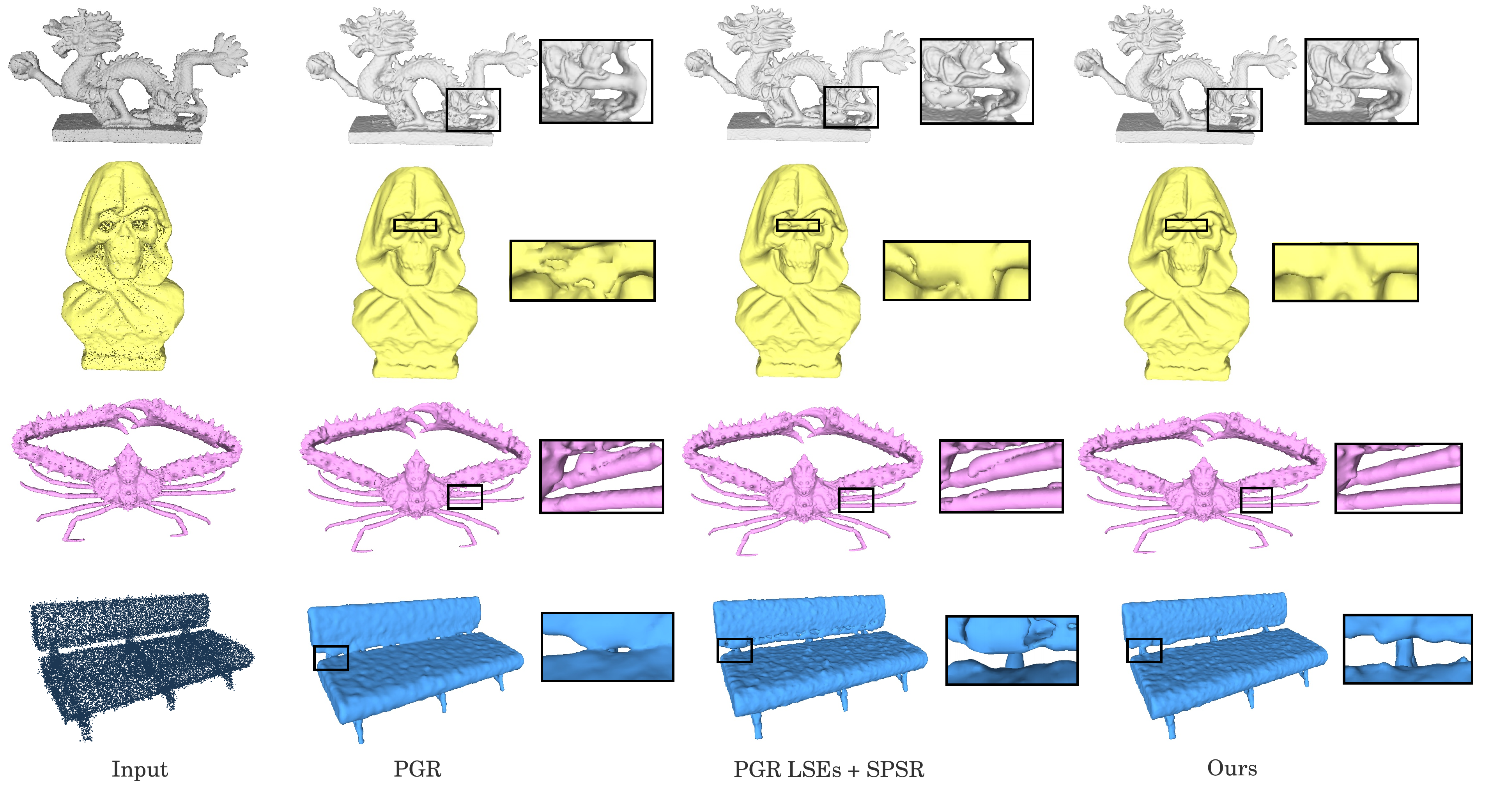}
  \caption{
           Qualitative comparisons with PGR~\citep{2022pgr}. PGR may over-smoothen the surface in noisy inputs. Our method performs better orientation accuracy and generates more details. }
    \label{fig:Figure4}
\end{figure}

\begin{figure}[htb]
  \centering
  \includegraphics[width=1.0\linewidth]{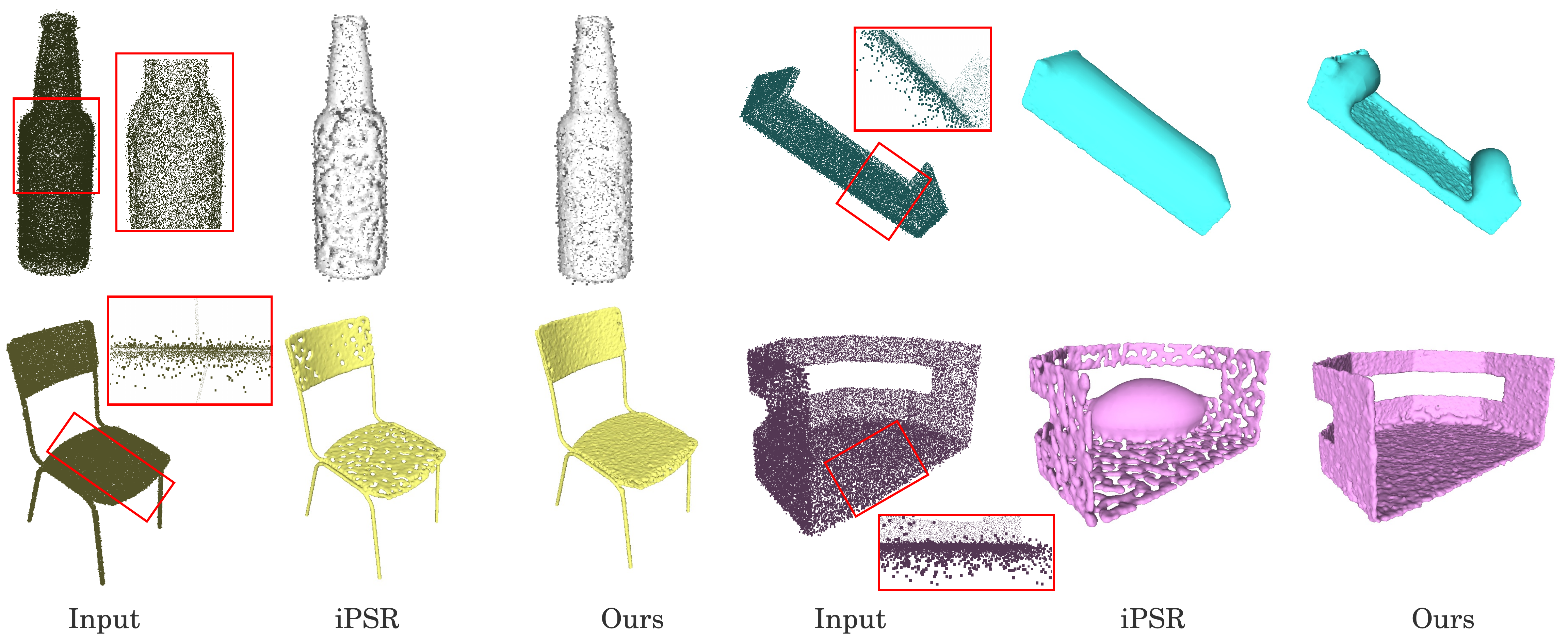}
  \caption{
           Qualitative comparisons with iPSR~\citep{2022ipsr}. Our method is good at identifying the normal orientation of the outermost surface and performs better in thin structures with noise.}
    \label{fig:Figure5}
\end{figure}

\begin{figure}[htb]
  \centering
  \includegraphics[width=0.95\linewidth]{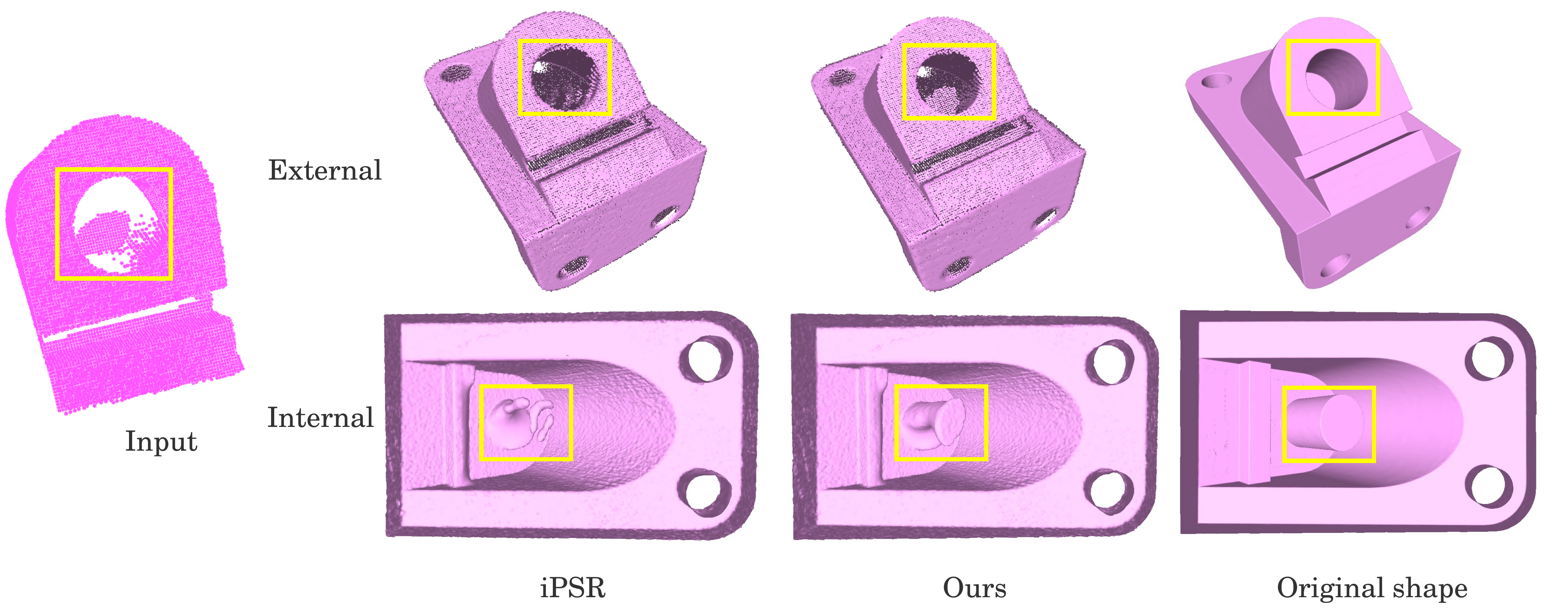}
  \caption{
           Results on the sampling with missing parts in the sunken cylinder of an anchor from both external and internal view. Our method achieves better orientation and reconstruction performance.}
    \label{fig:Figure6}
\end{figure}

\begin{figure}[htb]
  \centering
  \includegraphics[width=0.75\linewidth]{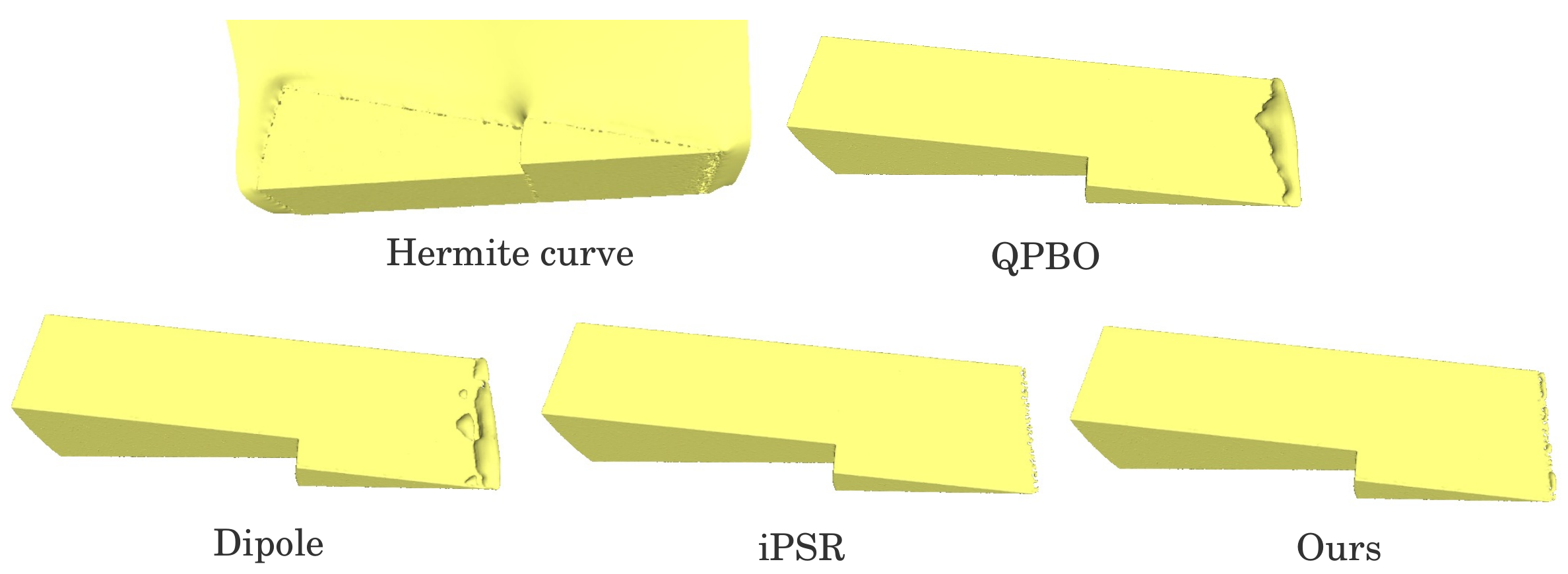}
  \caption{
         Results on sharp edges. All methods encounter some challenges to orient the sharp acute angle. Hermite curve even generates wrong orientation in the whole top side surface. Our approach achieves a relative decent effect in this situation.}
    \label{fig:Figure_sharp}
\end{figure}

\begin{figure}[htb]
  \centering
  \includegraphics[width=0.65\linewidth]{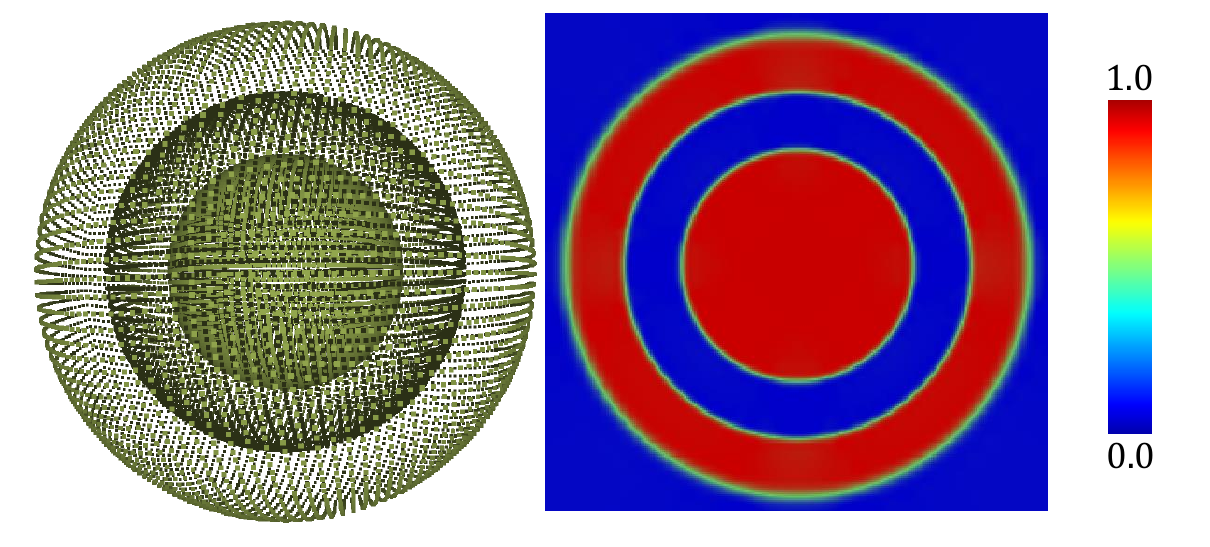}
  \caption{
           Our reconstruction on the nested surface with three layers on $r=0.5, r=0.75$, and $r=1.0$. We show the orientation results and the implicit space generated by the oriented point cloud. The green color represents the isovalue.}
    \label{fig:Figure_nested}
\end{figure}

\begin{figure}[htb]
  \centering
  \includegraphics[width=1.0\linewidth]{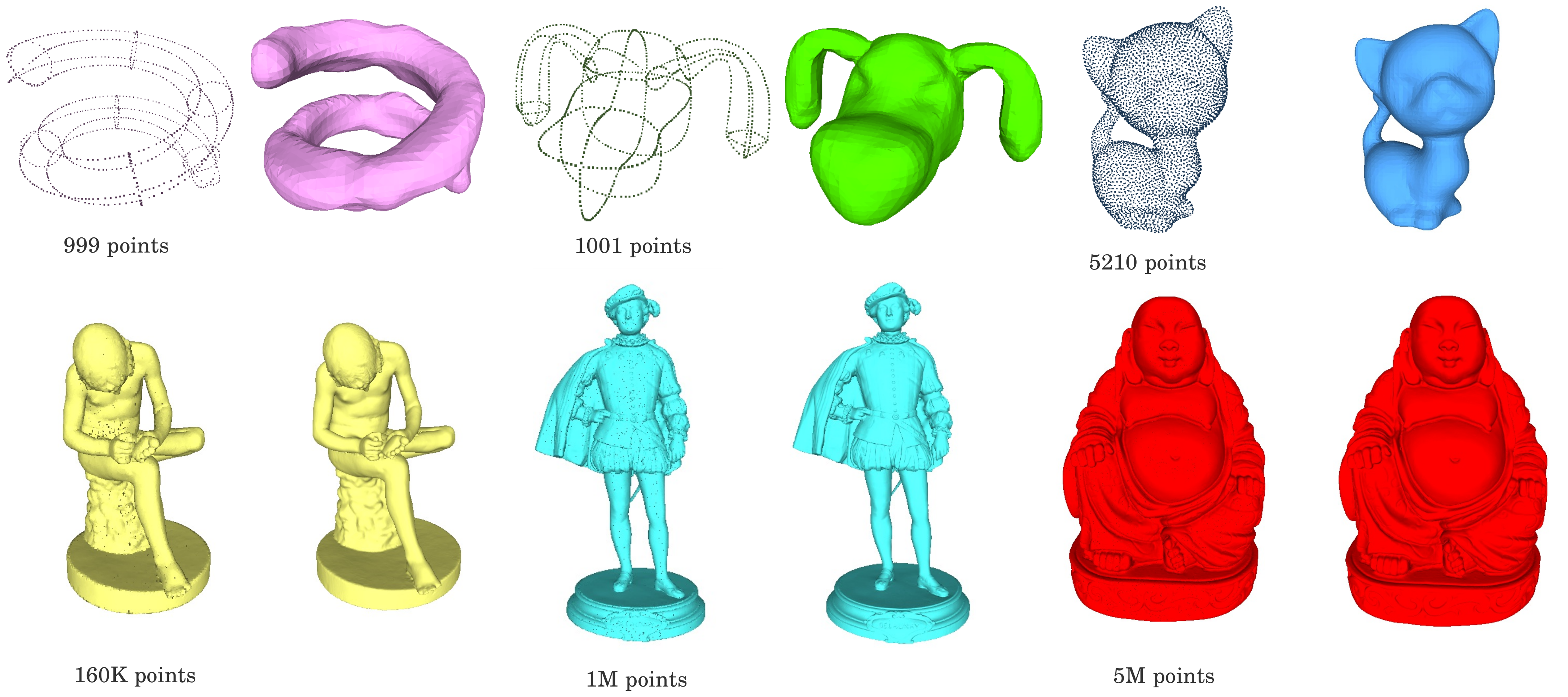}
  \caption{
           Our reconstructions on point clouds of different sampling densities from 3D sketch of 999 points to dense sampling with 5 million points.}
    \label{fig:Figure8}
\end{figure}

\begin{figure}[htb]
  \centering
  \includegraphics[width=1.0\linewidth]{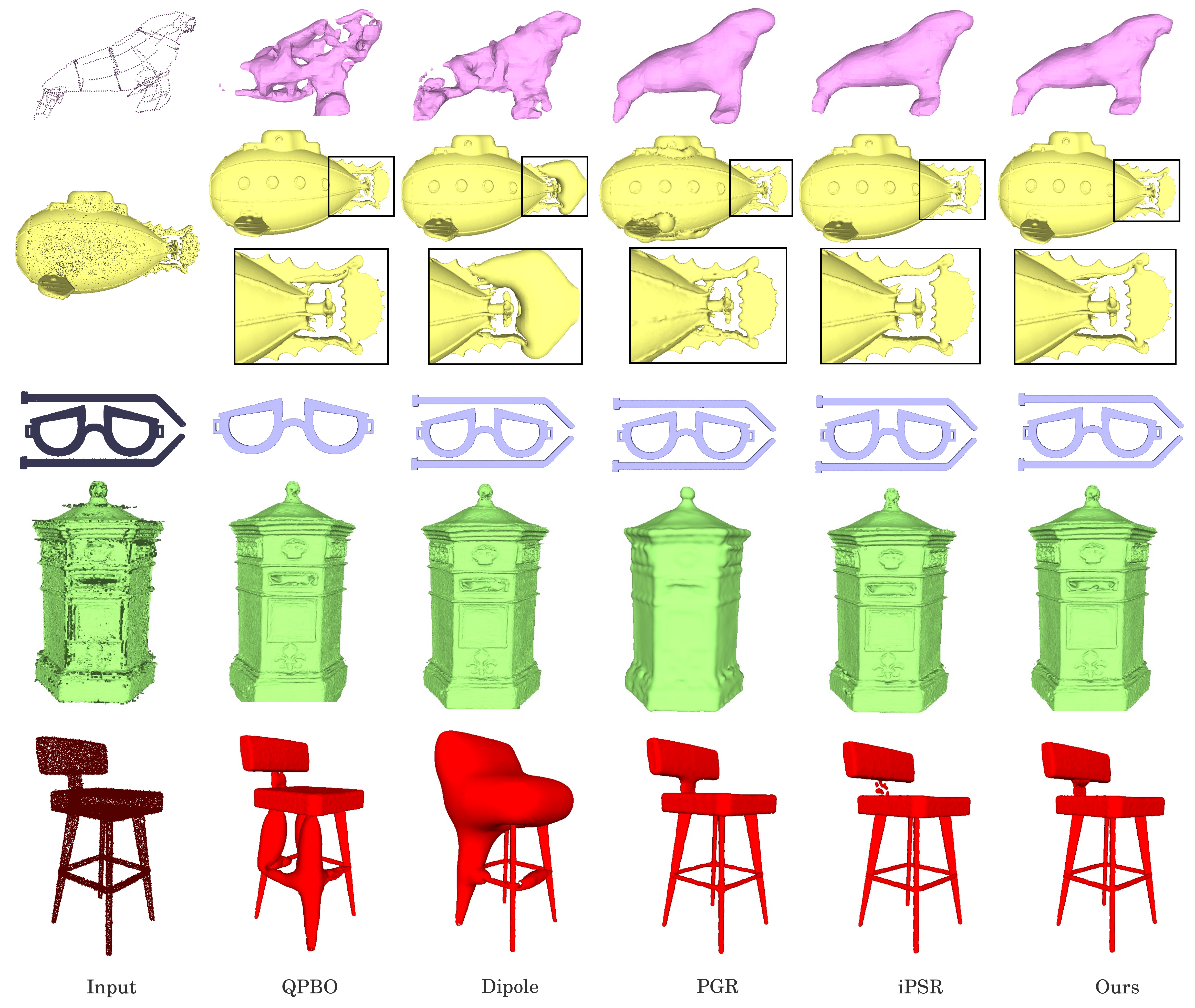}
  \caption{
           Qualitative comparisons on a variety of inputs involving extremely sparse samplings, complex topology, multiple connected components, outliers, and noise. Our method effectively manages these situations and achieves well-rounded performance among all methods.}
    \label{fig:Figure9}
\end{figure}

\begin{figure}[htb]
  \centering
  \includegraphics[width=1.0\linewidth]{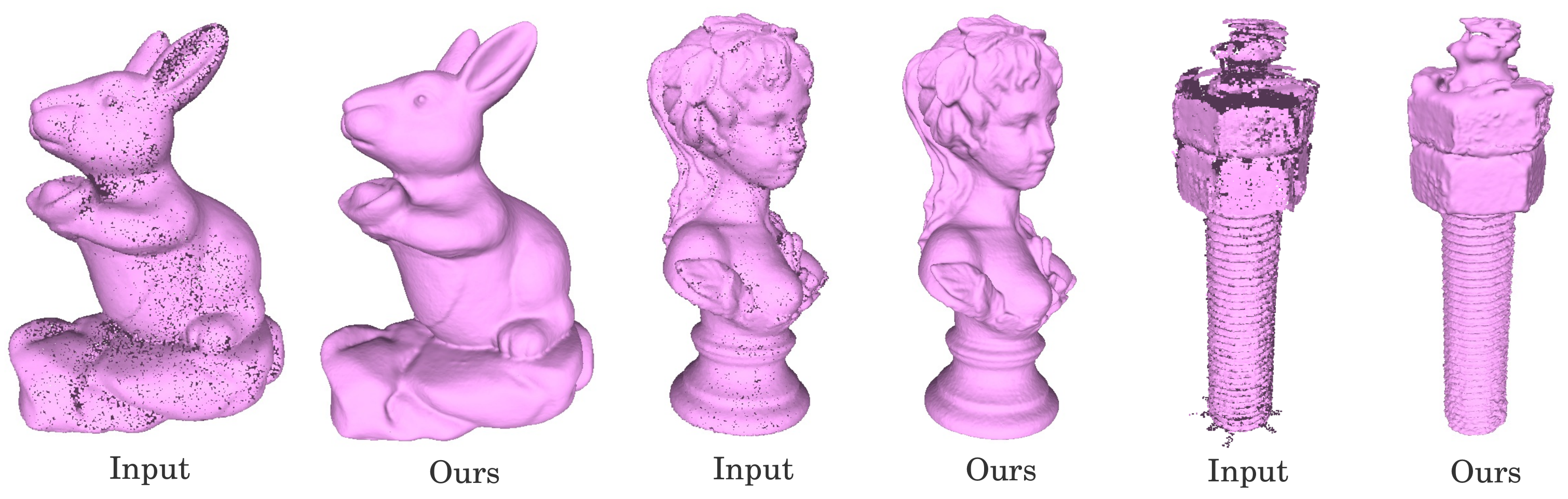}
  \caption{
           Our reconstructions on real scanned point clouds. }
    \label{fig:Figure_real}
\end{figure}

\begin{figure}[htb]
  \centering
  \includegraphics[width=0.75\linewidth]{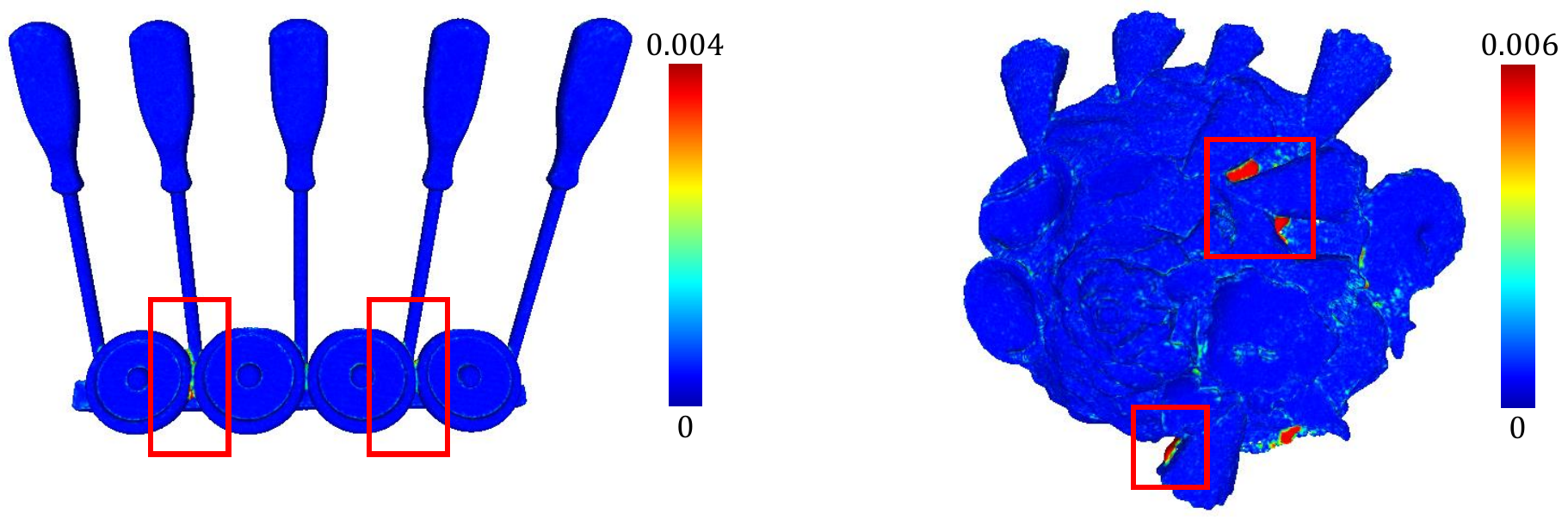}
  \caption{
           Our method may produce inaccurate orientation in non-outermost surfaces that are very close to each other. This is a limitation of our method.}
    \label{fig:Figure10}
\end{figure}

The qualitative comparisons are shown in Figure~\ref{fig:Figure3}. The oriented point clouds are visualized by MeshLab~\citep{2008Meshlab}. The first three examples are noise-free, and the last two contain thin structures with noise. Our method shows good orientation accuracy. MST, Hermite curve and QPBO are local consistency-based approaches. Therefore, local errors may cause large-scale inconsistent orientation. 
%%This phenomenon is also illustrated in the sharp angle of Figure~\ref{fig:Figure_sharp}.
Moreover, their performances degrade when the noise is large. Dipole separates the point clouds into patches, reducing its robustness when the wrong partition occurs, especially when thin structures exist. It is also sensitive to parameters and requires careful adjustment for complex shapes. Our approach exhibits high robustness for complex topology, thin structures, and noise.

If the L2 regularizer in Equation~\ref{equation_loc} is removed, then the orientation accuracy of our method on the ordinary var-noise dataset is reduced by $0.2\%$. Removing the entire $E_{loc}$ reduces the accuracy by $0.6\%$. This scenario indicates that our approach is dominated by isovalue constraints and achieves decent results without the local consistency energy. Meanwhile, the local consistency energy improves the average orientation accuracy of our method. We also visually demonstrate its effectiveness in Figure~\ref{fig:Figure_local}, where we show the reconstruction results with and without the local consistency energy $E_{loc}$ of two noisy inputs. We use black rectangles to emphasize the differences. It can be seen that $E_{loc}$ brings about better orientation consistency and improves the reconstruction quality near the zigzag region of the input shapes.

%%\subsection{Ablation Study}\label{sec:parameter}

In addition to orientation approaches, we also compare the reconstruction performance of our method with the most recent unoriented reconstruction approaches PGR~\citep{2022pgr} and iPSR~\citep{2022ipsr}. In iPSR (and SPSR), point weight is an important parameter that controls the degree of fitting points. We set the point weight to $10.0$ for noise-free inputs and $1.0$ for noisy inputs. In PGR, parameters $k_{w}$ and $\alpha$ control its ability to manage noise. We set $k_{w}=7, \alpha=2.0$ for noise-free inputs and $k_{w} = 64, \alpha = 5.0$ for noisy inputs.

PGR optimizes the product of normals and surface elements (also named as \textit{linearized surface elements} (LSEs) in~\cite{2022pgr}) in the Gauss formula. The linear system of PGR is dense. Therefore, it is always executed on GPU and limits to about 50K points in 2080Ti with 11GB memory. By contrast, our approach solves a sparse linear system and can work entirely on CPU. We can carry out per-point optimization of 100K points in common devices. Moreover, our method exhibits better average orientation accuracy in complex non-uniform and ordinary var-noise datasets. When normalized, the LSEs in PGR can also represent normals. The average orientation accuracy of PGR is $96.4\%$ in no-noise data and $91.9\%$ in var-noise data, and that of our approach is $99.3\%$ and $93.7\%$, respectively. In PGR, the surface elements are in the optimized variable. Moreover, several approximations are required to solve the singularity of the Gauss formula. Hence, the difference between LSEs and normals is evident. In Figure~\ref{fig:Figure4}, we show the reconstruction of PGR and the results of taking the normalized LSEs for orientation and reconstructing by SPSR. The results show that our method generates good details, for instance, the crab legs. PGR also over-smoothens the surface for noisy inputs.

iPSR iteratively processes SPSR and computes the normals from the surface of the preceding iteration. Although iPSR is an outstanding work and always performs high-fidelity reconstructions in complex topologies, its convergence currently has no theoretical proof due to the lack of explicit formulation of isosurfacing. Our method optimizes the normals in the implicit space and represents the solution by a linear equation. In Section~\ref{sec:overall}, we demonstrate that our method is good at identifying the outermost surface and the corresponding inward (outward) normals. Therefore, we can achieve good performance in thin structures with noise. This is shown in Figure~\ref{fig:Figure5} and the $5$th example of Figure~\ref{fig:Figure9}. In Figure~\ref{fig:Figure5}, the wall of the wine bottle and the base of the chair is thin with big noise. Therefore, judging whether the normal of a point should be inward the surface or outward is difficult. The results show that our approach deals with this situation better.

\begin{table}[t]
\centering
\caption{Running time of our approach on a usual laptop CPU. The time includes all IOs and reconstructions.}
\label{table2}
\begin{tabular}{ccccc}
\hline

Model  & Node count  & Point number & Representative set & Running time  \\
\hline

\textit{kitten} (3rd Figure~\ref{fig:Figure8}) & $67,753$ & $5,210$ & - & $23s$ \\
%\hline

\textit{bar\_chair36} (5th Figure~\ref{fig:Figure9}) & $155,953$ & $40K$ & - & $85s$ \\
%\hline

\textit{think10k66952} (2nd Figure~\ref{fig:Figure9}) & $167,761$ & $160K$ & $53K$ & $119s$ \\
%\hline

\textit{mailbox} (4th Figure~\ref{fig:Figure9}) & $209,929$ & $0.7M$ & $100K$ & $239s$ \\
%\hline

\textit{buddha} (6th Figure~\ref{fig:Figure8}) & $267,897$ & $5M$ & $100K$ & $456s$ \\
\hline

\end{tabular}
\label{table_MAP}
\end{table}

\subsection{Managing various situations}\label{sec:artifacts}
In this section, we examine the ability of our approach in handling various kinds of inputs, including artifacts not performed in Section~\ref{sec:compare} such as missing parts, sharp edges, and nested surfaces. We test our approach in different sampling densities from hundreds to millions. Moreover, we utilize real scanned point clouds to validate the effectiveness of our method.

Figure~\ref{fig:Figure6} shows an \textit{anchor} model with missing parts in the point cloud. The original shape of the anchor is shown in the rightmost column. The first row is the external view, and the second row is the internal view (looking from the internal of the shape). However, the sunken cylinder of the anchor is not scanned completely, causing a missing region in the input. The left two columns are orientation and reconstruction performances of iPSR and ours with the scanned data. It can be seen from the external view that our method generates correct normal orientation for the sunken cylinder although only a few points are scanned in this region. Therefore, we achieve better reconstruction performance which is obvious from the internal view.

Figure~\ref{fig:Figure_sharp} shows a comparison of sharp edges among different approaches. We can see that all methods encounter some challenges to orient the sharp acute angle. Hermite curve even generates wrong orientation in the whole top side surface. Our approach achieves a relative decent effect in this situation. In Figure~\ref{fig:Figure_nested}, we exhibit an example of nested surface with three layers at $r=0.5, r=0.75$, and $r=1.0$. We show the orientation result and the implicit space generated by the oriented point cloud. The green color represents the isovalue. Our method generates correct normal orientation: the normals are alternative in and out among adjacent layers. Managing nested surfaces is also one of the advantages of our method over visibility-based and propagation-based approaches. 

Figure~\ref{fig:Figure8} shows our reconstructions of different sampling densities from 3D sketch of $999$ points to dense sampling with $5$ million points. Our method exhibits high performance for these examples. In Figure~\ref{fig:Figure9}, we show qualitative comparisons on a variety of inputs involving extremely sparse samplings, complex topology, multiple connected components, outliers, and noise. It can be seen that our method effectively manages these situations and achieves well-rounded performance among all methods. In Figure~\ref{fig:Figure_real}, we show our reconstructions of several real scanned point clouds provided by~\cite{2022Survey}. Our method produces consistently oriented normals and therefore generates detailed reconstruction when the scanned point cloud is accurate. Even when the scanned point cloud is noisy with missing regions, our method can still achieve a decent performance.

\subsection{Running time}\label{sec:time}
Table~\ref{table2} exhibits the running time of our approach in a laptop with AMD Ryzen 5 5600H CPU @ 3.3GHz and total of $24$GB memory in two memory slots. The running time includes all processes: IOs, orientations, reconstructions, representative point samplings, etc. We show the tree node count (during optimization), point number, representative set point number, and running time of several examples. If the representative set point number is ``-'' in the table, then we have conducted per-point optimization (no representative set is used) in this shape. For the \textit{buddha} model with $5$ million points, we sample $100$K points for the representative set. The time for orienting representative points is $214$s, and the time for other processes (all IOs, orienting the dense point cloud by the representative set) is $242$s, for a total of $456$s. For other examples, most time is spent on optimization.

We take \textit{bar\_chair36} with 40K points for comparison among different methods. The running time includes orientation and reconstruction. We do not apply representative set in this example and records $85$s. Local consistency-based methods MST, Hermite curve and QPBO are fast in CPU and only take $4$s, $12$s and $4$s respectively. iPSR spends $49$s in $17$ iterations. Dipole applies a deep neural network to keep local orientation consistency. Therefore, it relies on GPU and takes $211$s on a server with Intel(R) Xeon(R) Silver 4210 CPU @ $2.20$GHz and RTX $2080$Ti GPU. PGR solves a dense linear system and relies heavily on computing resources with time complexity quadratic to the point numbers. We change its kdtree implementation from ``KDTree'' to ``cKDTree'' in SciPy~\citep{DBLP:journals/corr/abs-1907-10121} to achieve fast width calculation. It takes about $87$s on the RTX 2080Ti GPU. If executed in CPU, then it takes several hours. Applying a representative set reduces the running time of our method. For example, our approach takes $239$s for \textit{mailbox}, and iPSR takes $854$s in $30$ iterations.

\section{Conclusion and future works}\label{sec:conclusion}
In this work, we propose an orientation and reconstruction approach that incorporates isovalue constraints to the Poisson equation. We bridge orientation and reconstruction in the implicit space and express the orientation problem into a sparse linear system. Our method shows competitive performance for varying noise and artifacts, and works on an average laptop CPU. We believe that communicating orientation and reconstruction in the implicit function space is interesting and follow-up works can be carried out.

Our method still has some limitations. Although our approach does not rely on GPU, the solving process of the linear system still requires a large amount of memory (about $10.9$GB for \textit{mailbox} and $12.6$GB for \textit{buddha}). It mainly depends on the tree node count. However, most laptops have more than one memory slots to support the usage of $16$GB to even $32$GB. In the future, we can try the depth-by-depth solving similar to SPSR instead of including all nodes in a large linear system. Moreover, our method may produce inaccurate orientation in non-outermost surfaces that are very close to each other. Figure~\ref{fig:Figure10} shows the error map of our reconstructions relative to the ground truth meshes. The high error places demonstrate this limitation. The possible reason is that the indicator function changes rapidly in this area. However, the octree depth during optimization is not sufficiently high (recall that the implicit space is spanned by B-spline basis functions centered at octnodes). Conducting depth-by-depth solving is still a potential solution to support large depth during optimization.

\appendix
\renewcommand{\appendixname}{Appendix}

\section{Details of the conjugate gradient optimization}\label{appendixA}

This appendix presents the details of the conjugate gradient optimization of our method. It also explains why the sparsity of the matrices can be maintained during the solving process. In Section~\ref{sec:solving}, the optimization problem is transformed into a solution of a linear system
\begin{equation}
\tilde{A}\tilde{x} = \tilde{b},
\end{equation}
where
\begin{equation}
\tilde{A}=
\begin{pmatrix}
    U^T U + \alpha A^T A & -\alpha A^T B\\
    -\alpha B^T A & \alpha B^T B + \beta M\\
\end{pmatrix},
\qquad
  \tilde{x} =
\begin{pmatrix}
    x\\
    n\\
 \end{pmatrix},
 \qquad
\tilde{b} =
\begin{pmatrix}
    U^T \frac{1}{2} \vec{1}\\
    0\\
\end{pmatrix}.
\end{equation}

Let ${\tilde{x}^{(k)}} = (x^{(k)}, n^{(k)})$ be the results of the $k$-th iteration and is initialized as $x^{(0)} = 1e^{-3} \times \vec{1}, n^{(0)} = \vec{0}$. The conjugate graident algorithm can be written as follows:
let ${\tilde{r}}^{(0)} = \tilde{b} - \tilde{A}\tilde{x}^{(0)}, {\tilde{p}}^{(0)} = {\tilde{r}}^{(0)}$, 
where
\begin{equation}
\tilde{A}\tilde{x}^{(0)} =
\begin{pmatrix}
   U^T U x^{(0)} + \alpha A^T A x^{(0)} - \alpha A^T B n^{(0)}\\
   - \alpha B^T A x^{(0)} + \alpha B^T B n^{(0)} + \beta M n^{(0)}\\
\end{pmatrix}.
\end{equation}
Denote
\begin{equation}
  \tilde{p}^{(k)} =
\begin{pmatrix}
    p_{x}^{(k)}\\
    p_{n}^{(k)}\\
 \end{pmatrix},
 \qquad
\tilde{r}^{(k)} =
\begin{pmatrix}
  r_{x}^{(k)}\\
  r_{n}^{(k)}\\
 \end{pmatrix},
\end{equation}
then, for each iteration,
\begin{align}
  & \alpha_{k} = \frac{(\tilde{r}^{(k)}, \tilde{r}^{(k)})}{(\tilde{p}^{(k)}, \tilde{A}\tilde{p}^{(k)})} = \frac{(r_{x}^{(k)}, r_{x}^{(k)}) + (r_{n}^{(k)}, r_{n}^{(k)})}{(U p_{x}^{(k)}, U p_{x}^{(k)}) + \alpha [(A p_{x}^{(k)}, A p_{x}^{(k)}) + (B p_{n}^{(k)}, B p_{n}^{(k)}) - 2 (A p_{x}^{(k)}, B p_{n}^{(k)})] + \beta (p_{n}^{(k)}, M p_{n}^{(k)})}, \\
 & \tilde{x}^{(k+1)} = \tilde{x}^{(k)} + \alpha_{k} \tilde{p}^{(k)}, \\
  & \tilde{r}^{(k+1)} = \tilde{r}^{(k)} - \alpha_{k} \tilde{A} \tilde{p}^{(k)}, \\
& \beta_{k} = \frac{(\tilde{r}^{(k+1)},  \tilde{r}^{(k+1)})}{(\tilde{r}^{(k)}, \tilde{r}^{(k)})} = \frac{(r_{x}^{(k+1)}, r_{x}^{(k+1)}) + (r_{n}^{(k+1)}, r_{n}^{(k+1)})}{(r_{x}^{(k)}, r_{x}^{(k)}) + (r_{n}^{(k)}, r_{n}^{(k)})}, \\
& \tilde{p}^{(k+1)} = \tilde{r}^{(k+1)} + \beta_{k} \tilde{p}^{(k)},
  \end{align}
  where
\begin{equation}
\tilde{A} \tilde{p}^{(k)} =
\begin{pmatrix}
   U^T U p_{x}^{(k)} + \alpha A^T A p_{x}^{(k)} - \alpha A^T B p_{n}^{(k)}\\
   - \alpha B^T A p_{x}^{(k)} + \alpha B^T B p_{n}^{(k)} + \beta M p_{n}^{(k)}\\
\end{pmatrix}.
\end{equation}

Note that $U^T U x = (U^T(Ux))$ and $x^TU^TUx=(Ux, Ux)$. Therefore, each step of the conjugate gradient algorithm can be written as the product of a matrix and a vector, or the inner product of two vectors. The sparsity of the matrices can then be maintained during the solving process. The complexity and the number of iterations are described in Section~\ref{sec:solving}.
%%%%%%%%%%%%%%%%%%%%%%%%%%%%%%%%%%%%%%%%%%%%%%%%%%%%%%%%%%%%%%%%%%%%%
%\section*{References} % needed on some systems
\bibliography{mybibfile}

\end{document}